\documentclass[pre,aps,twocolumn,english,showpacs,superscriptaddress]{revtex4-1}
\usepackage[T1]{fontenc}
\usepackage[latin9]{inputenc}
\usepackage{color}
\usepackage{amsmath}
\usepackage{graphicx}
\usepackage[normalem]{ulem}
\usepackage{hyperref}
\usepackage{cleveref}

\usepackage{babel}

\usepackage{ulem}
\renewcommand{\Re}{\operatorname{Re}}

\begin{document}
\author{A.~G.~Vladimirov}
\email{vladimir@wias-berlin.de}
\affiliation{Weierstrass Institute for Applied Analysis and Stochastics, Mohrenstrasse 39, D-10117 Berlin, Germany}
\author{A.~V.~Kovalev}
\email{antony.kovalev@gmail.com}
\affiliation{ITMO University,  14 Birzhevaya line, Saint-Petersburg, 199034, Russia}
\author{E.~A.~Viktorov}
\email{evviktor@gmail.com}
\affiliation{ITMO University,  14 Birzhevaya line, Saint-Petersburg, 199034, Russia}
\author{N.~Rebrova}
\email{nrebrova@gmail.com}
\affiliation{Department of Physical Sciences, Cork Institute of Technology, Cork, Ireland}
\author{G.~Huyet}
\email{guillaume.huyet@inphyni.cnrs.fr}
\affiliation{Universit\'e C\^ote d'Azur, CNRS, INPHYNI, France}

\title{Dynamics of a class A nonlinear mirror mode-locked laser}          

\begin{abstract}
Using a delay differential equation model we study theoretically the dynamics of a unidirectional class-A ring laser with a nonlinear amplifying loop mirror.  We perform linear stability analysis of the CW regimes in the large delay limit and demonstrate that these regimes can be destabilized via modulational and Turing-type instabilities, as well as by an instability leading to the appearance of square-waves. We investigate the formation of square-waves and mode-locked pulses in the system.  We show that mode-locked pulses are asymmetric with exponential decay of the trailing edge in positive time and faster-than-exponential (super-exponential) decay of the leading edge in negative time. We discuss asymmetric interaction of these pulses leading to a formation of harmonic mode-locked regimes. 
 \end{abstract}
\maketitle

\section{Introduction}
\label{sec:intro}  

The possibility of generation of short light pulses by locking the
longitudinal modes of a laser was discussed only a few years after the 
development of the laser in 1960. Mode-locking techniques can be classified
into two major classes: (i) active mode-locking, based on an external
modulation at a frequency close to the cavity free spectral range and (ii)
passive mode-locking where an intracavity nonlinear component reduces losses
for pulsed operation with respect to the those of continuous-wave (CW)
regime. A standard theoretical approach to study  the properties of
mode-locked devices is based on direct integration of the so-called
{\em traveling wave equations} describing space-time evolution of the electric
field and carrier density in the laser sections \cite%
{Tromborg94,Avrutin00,Bandelow01,Bandelow06}. Another, much simpler approach
limited to small gain and loss approximation was developed by Haus in \cite{Haus00}.
To overcome this limitation the third modelling approach was suggested in 
\cite{VT04,Vladimirov,VT05} based on the lumped element method that allows
to derive a delay differential equation (DDE) model for the temporal
evolution of the optical field at some fixed position in the cavity. DDE
models successfully describe the dynamics of multi-section mode-locked
semiconductor lasers \cite%
{VMVB06,RVBHK06,vladimirov2010dynamical,arkhipov,Marconi,ViktorovCoherence},
frequency swept light sources \cite{Slepneva,Slepneva2,pimenovprl},
optically injected lasers \cite{rebrova,pimenov2014}, semiconductor lasers
with feedback \cite{Kelleher,otto,Jaurigue17}, as well as some other
multimode laser devices \cite{ViktorovOL}. 

In this work, we propose and study a DDE model for nonlinear amplifying loop mirror (NALM) mode-locked laser. 
Nonlinear mirror laser as a device for ultrafast light processing was proposed in \cite{doran1988nonlinear}. 
Mode-locked pulse formation mechanism in this laser is based on 
the asymmetric nonlinear propagation in a waveguide loop, where two counterpropagating waves aquire  different intensity-dependent phase shifts caused by the Kerr nonlinearity. 
As a result of the interference of these waves the loop acts as a nonlinear mirror with the reflectivity dependent on the intensity of incident light. Such nonlinear mirror plays a role of a saturable absorber that leads to the appearance of a pulsed laser operation.

We develop a DDE laser model by assuming dispersion-free unidirectional operation inside
the cavity and symmetrical beam splitting of the field into two
counter-propagating fields at the entrance of the NALM.  When the material variables are adiabatically eliminated, one obtains a single DDE for the complex electric field envelope, which, despite of its simplicity, gives a good insight on the laser dynamics.  Using this equation we find different mode-locked regimes of laser operation including square waves, ultrashort optical pulses and their harmonics. The mode-locked pulses are always bistable with the non-lasing state, and, hence, can also be considered as
temporal cavity solitons or nonlinear localized structures of light \cite{leo2010temporal,Marconi}. The linear stability analysis of CW solutions corresponding to different longitudinal modes of the laser reveals modulational and Turing instabilities, as well as an instability leading to the emergence of square waves. 
We demonstrate analytically the mode-locked pulses are asymmetric with exponential decay of the pulse trailing tail in positive time and faster-than-exponential (super-exponential) decay of the leading tail in  negative time. We find that the repulsive interaction of such asymmetric pulses leads to a harmonic mode-locking regime. In order to explain the experimentally observed square wave generation recently reported in \cite{Aadhi} in the figure-of-eight laser, we construct a one-dimensional map exhibiting a period doubling route to chaos.

\section{Model equation and CW solutions}
\begin{figure}[tbp]
\includegraphics[width=0.45\textwidth]{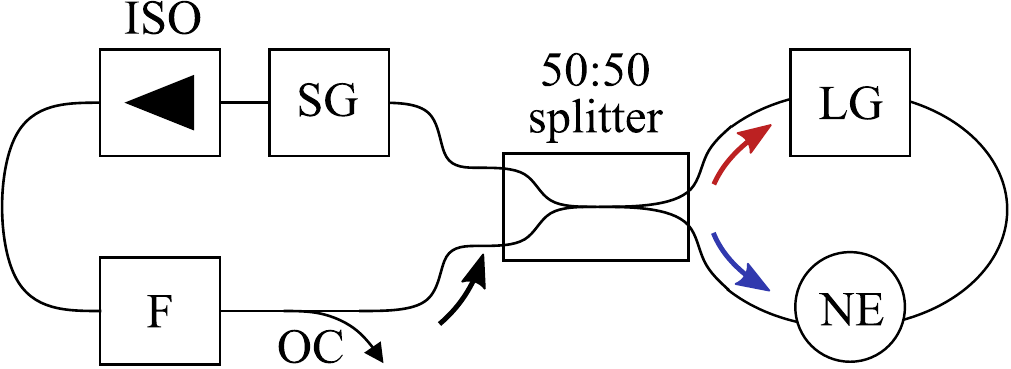} \centering
\caption{Schematic view of a ring laser with a saturable gain (SG) medum, bandpass filter (F), optical isolator (ISO), output coupler (OC), and linear gain (LG) together with a
non-linear element (NE) in a Sagnac interferometer forming a NALM.
Arrows show different propagation directions inside the
interferometer. }
\label{fig:scheme}
\end{figure}

A schematic of a 
NALM laser with a gain  and a spectral filter in an unidirectional cavity coupled to a bidirectional loop  with a second gain medium and a nonlinear element, is shown in Fig.~\ref{fig:scheme}. This scheme corresponds to experimentally implemented setups of mode-locked lasers with a high-Q microring resonator \cite{Kues2017} and an integrated waveguide \cite{Aadhi} acting as nonlinear elements. Using the approach of \cite{Vladimirov,VT05} we write the following DDE model for the time evolution of the electric field amplitude $E$ and saturable gain $g$ in a  laser shown in Fig.~\ref{fig:scheme}: 
\begin{equation}
\gamma^{-1}\frac{dE}{dt}+E=\sqrt{\kappa_1}e^{g/2+i\theta}{\tilde f}\left[|E(t-T)|^{2},G\right]E(t-T),
\label{eq:A}
\end{equation}
\begin{equation}
\gamma_{g}^{-1}\frac{dg}{dt}=g_{0}-g-\left(e^{g}-1\right)|E(t-T)|^{2}.
\label{eq:B}
\end{equation}
Here $\gamma $ is the spectral filter bandwidth, $\kappa_1$ is the linear
attenuation factor describing nonresonant cavity losses in the unidirectional part of the laser cavity, $g_{0}$ is the pump parameter, $\gamma_g$ is the relaxation rate of the amplifying medium, $\theta$ describes the detuning between the central frequency of the filter and one of the cavity modes, and the delay parameter $T$ is equal to the cold cavity round trip time.

The function ${\tilde f}\left(|E|^2,G\right)$  in Eq.~\eqref{eq:A} describes the reflectivity and the phase shift introduced by the NALM. It is given by ${\tilde f}\left(|E|^2,G\right)=\sqrt{\kappa_2}e^{G/2}\left(e^{i\phi_1}-e^{i\phi_2}\right)/2$, where  $G$ and  $\kappa_2$ and describe amplification and linear losses inside the loop, while $\phi_1=\eta |E|^2/2$  and $\phi_2=\eta e^G|E|^2/2$ are the phase shifts of the clockwise and counter-clockwise propagating waves  indicated by arrows in Fig.~\ref{fig:scheme}. Since the counter-clockwise propagating wave is amplified before passing through the nonlinear element its phase shift $\phi_2$  is $e^G$ times larger than that the clockwise propagating wave, which is amplified after passing through the nonlinear element.  For the gain $G$ an equation similar to Eq.~\eqref{eq:B} can be written. Here, however,  we assume for simplicity that the electric field intensity $|E|^2$ is small and gain medium inside the NALM operates far below the saturation regime.  In this case $G$ can be considered to be a constant parameter. Furthermore, for $\phi_1=\eta |E|^2/2\ll1$ we have $e^{\phi_1}\approx 1$  and the function ${\tilde f}$ can be rewritten in the form
\begin{equation*}
{\tilde f}(|E|^{2},G)=\sqrt{\kappa_2}e^{G/2}f(|E|^{2}),
\end{equation*}%
with
\begin{equation}
f(|E|^{2})=\frac{1}{2}\left(1-e^{ia|E|^{2}}\right), \label{eq:f}
\end{equation}
where $a=\eta (e^G-1)/2$. Finally, replacing $e^g-1$ with $g$ in Eq.~\eqref{eq:B}, eliminating the gain $g$ adiabatically, $g=g_0/\left[1+|E(t-T)|^2\right]$, and substituting this expression into Eq.~\eqref{eq:A} we get
\begin{equation}
\gamma ^{-1}\frac{dE}{dt}+E=\sqrt{\kappa }%
e^{g_{0}/[2(1+|E(t-T)|^{2})]+i\theta}f(|E(t-T)|^{2})E(t-T), \label{eq:model}
\end{equation}%
where $\kappa=\kappa_1\kappa_2e^{G/2}<1$, which means that the linear gain in the NALM is compensated by the linear cavity losses, and the nonlinear function $f(|E|^2)$ is defined by Eq.~\eqref{eq:f}. In the following
we will show that despite being very simple, the model equation (\ref%
{eq:model}) is capable of reproducing such experimentally observed
behaviours of nonlinear mirror lasers as mode-locking and square wave generation \cite{Aadhi}.

\begin{figure*}
\includegraphics[width=0.45\textwidth]{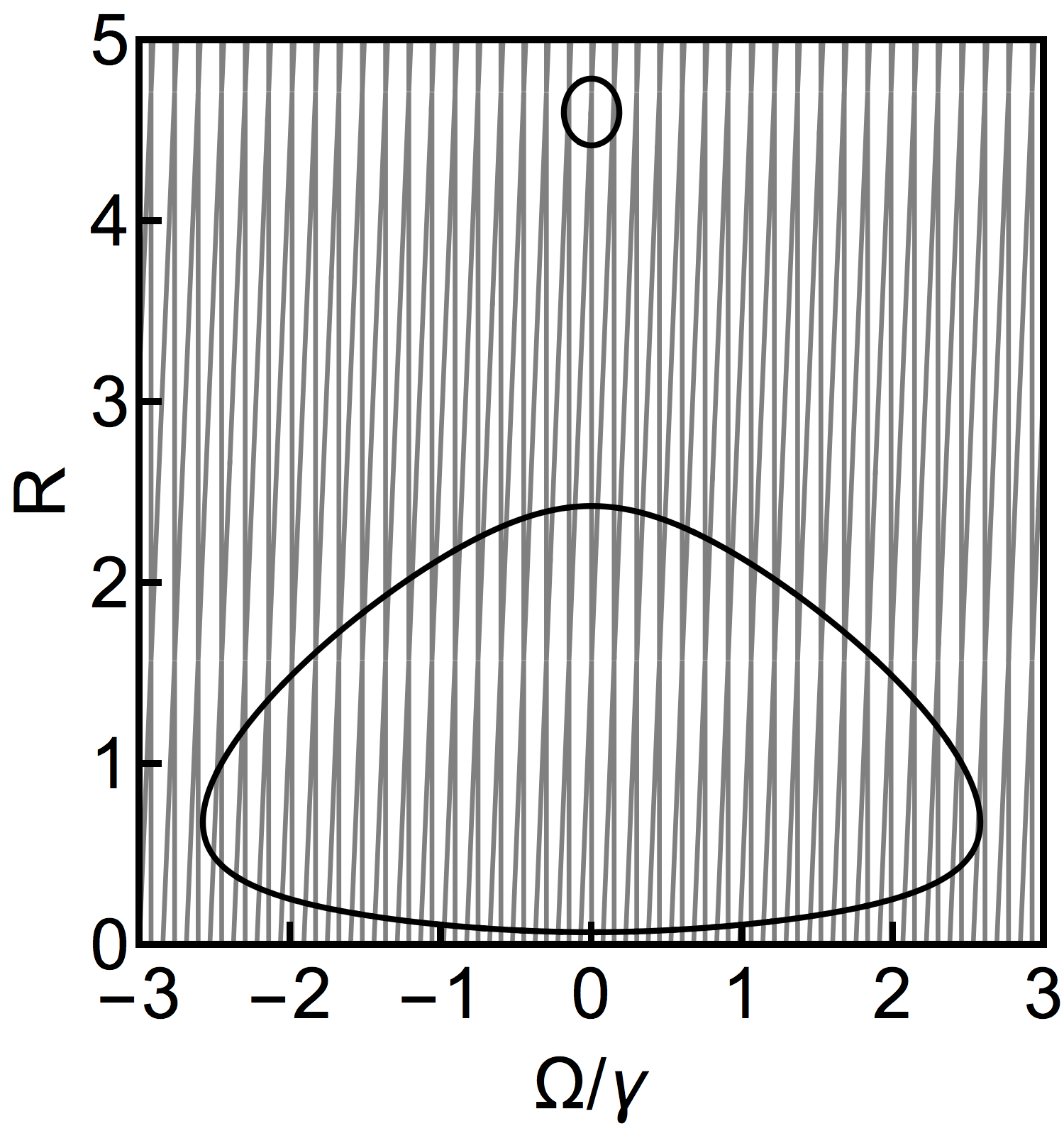}\quad
\includegraphics[width=0.45\textwidth]{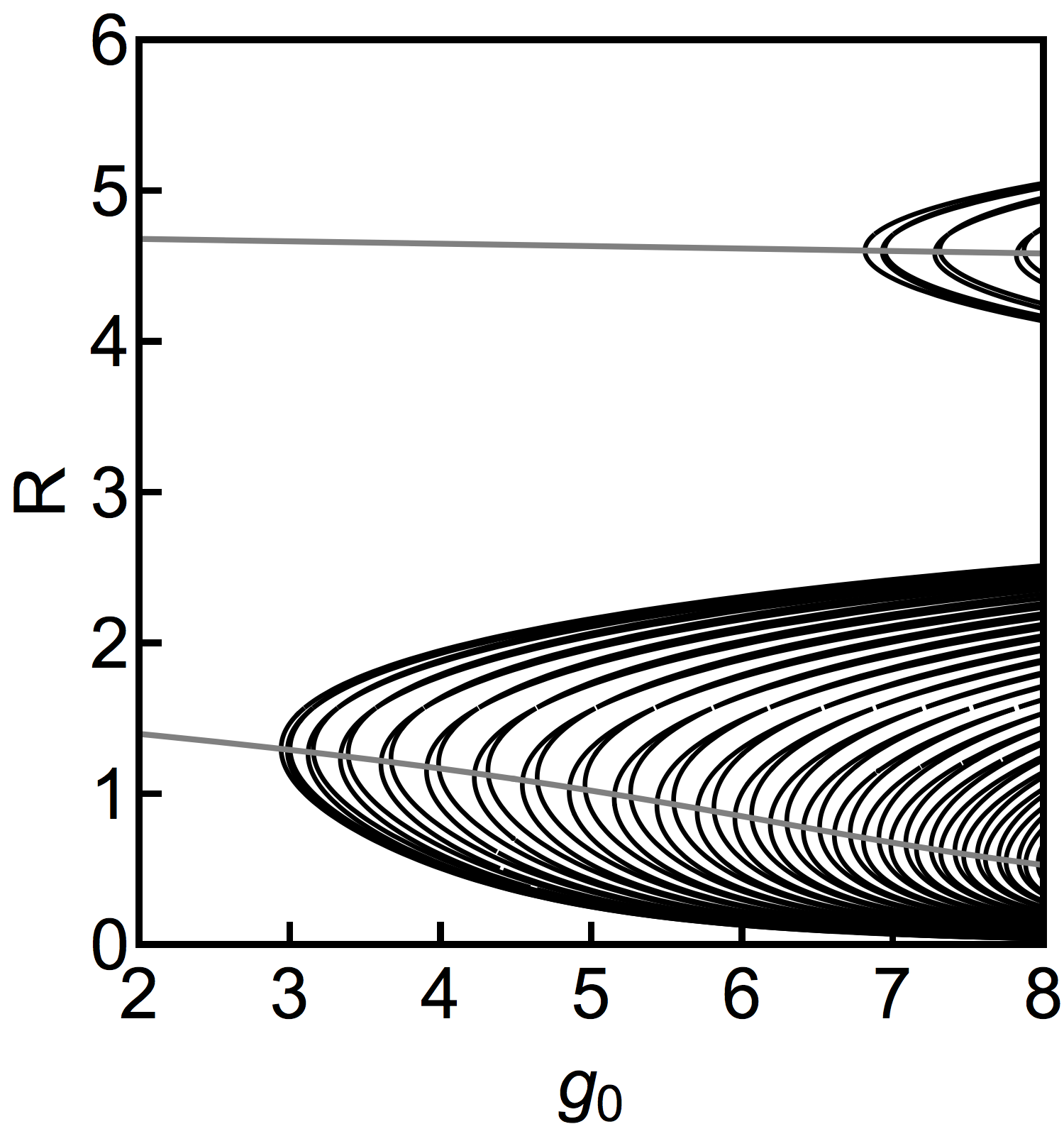}
\centering
\caption{Left: CW solutions at fixed pump parameter $g_0=5.0$.  CW solutions lie on the  intersections of the closed black curves [solutions of Eq.~(\ref{eq:CW1})] with thin gray lines [solutions of Eq.~(\ref{eq:CW2})].  
Other parameter values: $a=2$, $\gamma=1$, $\kappa=0.3$, $\theta=0$, and $T=20$.
Right: Branches of CW solutions  corresponding to different longitudinal
laser modes (black lines). Two gray lines are defined by the condition (\ref{eq:SN}). They intersect CW branches at the fold bifurcation points. Lower parts of the CW branches lying below the fold bifurcation points are always unstable, while upper parts can be either stable or unstable. 
\label{fig:CWbranches}}
\end{figure*}

The simplest solution of Eq.~(\ref{eq:model}) is that corresponding to laser
off state, $E=0$. Linear stability analysis of this trivial non-lasing
solution indicates that it is always stable, which means that the laser is
non-self-starting for all possible parameter values. Non-trivial CW
solutions are defined by the relation 
\begin{equation}
E=\sqrt{R}e^{i\Omega t},\label{eq:CW}
\end{equation}
where $R=|E|^{2}>0$ is the intensity and $\Omega $ is the frequency detuning of CW
regime from the reference frequency coinciding with the central frequency of
the spectral filter. $R$ and $\Omega$ are the solutions of a system of two transcendental equations 
\begin{equation}
\kappa e^{\frac{g_{0}}{1+R}}\sin ^{2}\left( \frac{aR}{2}\right) =1+\frac{%
\Omega ^{2}}{\gamma ^{2}},  \label{eq:CW1}
\end{equation}%
\begin{equation}
\tan \left( \frac{aR^{2}}{2}-T\Omega +\theta\right) +\frac{\gamma }{\Omega }=0.
\label{eq:CW2}
\end{equation}%
Multiple solutions of these equations corresponding to different longitudinal modes of the laser are illustrated in the left panel of
Fig~\ref{fig:CWbranches}, where CW regimes correspond to the intersections of black closed curves obtained by solving Eq.~\eqref{eq:CW1} and thin gray lines calculated from Eq.~\eqref{eq:CW2}. Black curves in right panel of Fig.~\ref{fig:CWbranches} show the branches of non-trivial CW
solutions as functions of the pump parameter $g_{0}$, while gray lines  are defined by the condition
\begin{equation}
g_{0}=\tilde{g}\equiv a(1+R)^{2}\cot \left( \frac{aR}{2}\right). 
\label{eq:SN}
\end{equation}%
The intersections of the gray lines with black CW branches indicate fold bifurcation points, each of which separates the corresponding branch into two parts with the lower part being always unstable.  

\section{CW  stability in the large delay limit}
\begin{figure*}
\includegraphics[width=0.45\textwidth]{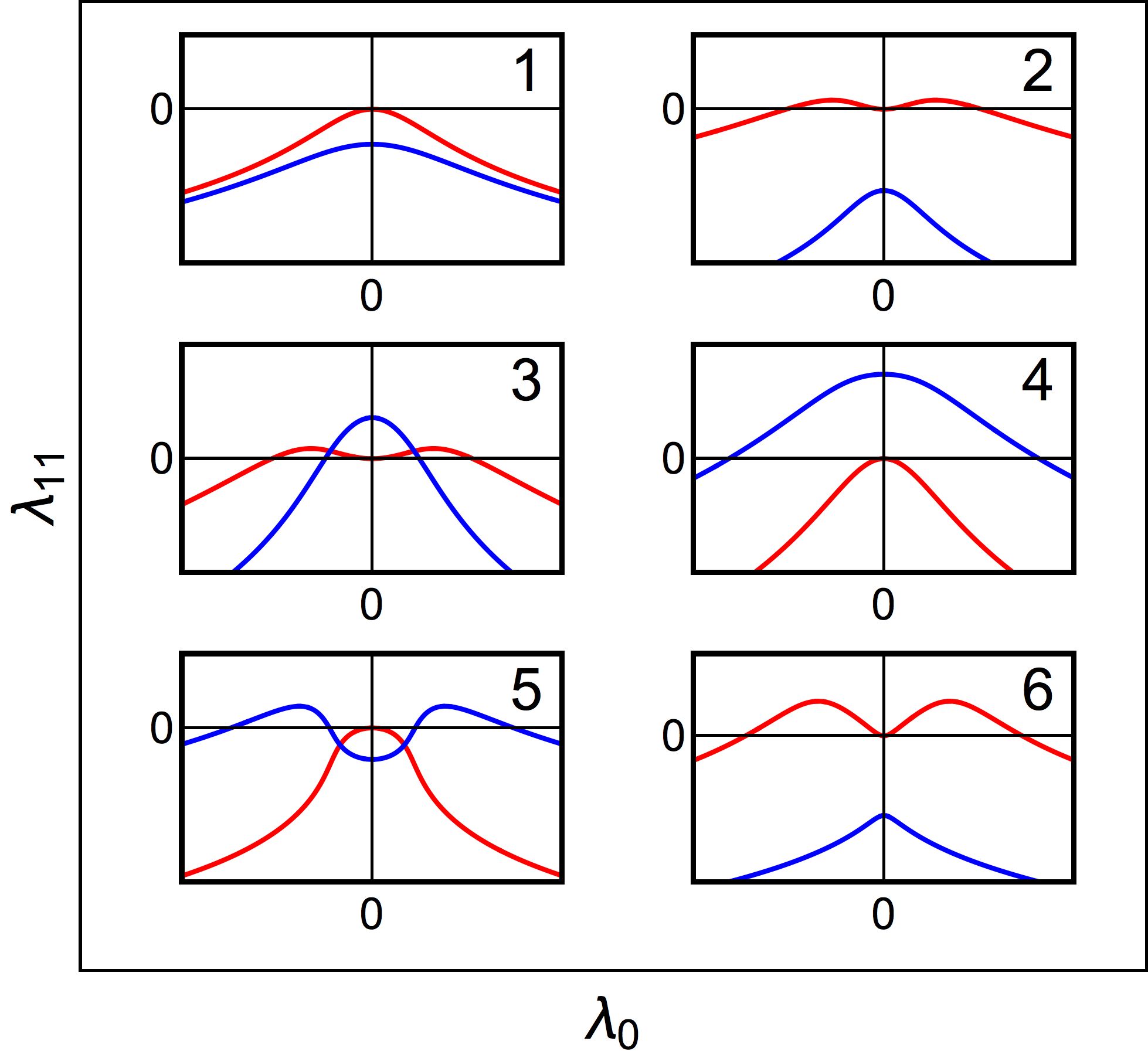}\quad
\includegraphics[width=0.45\textwidth]{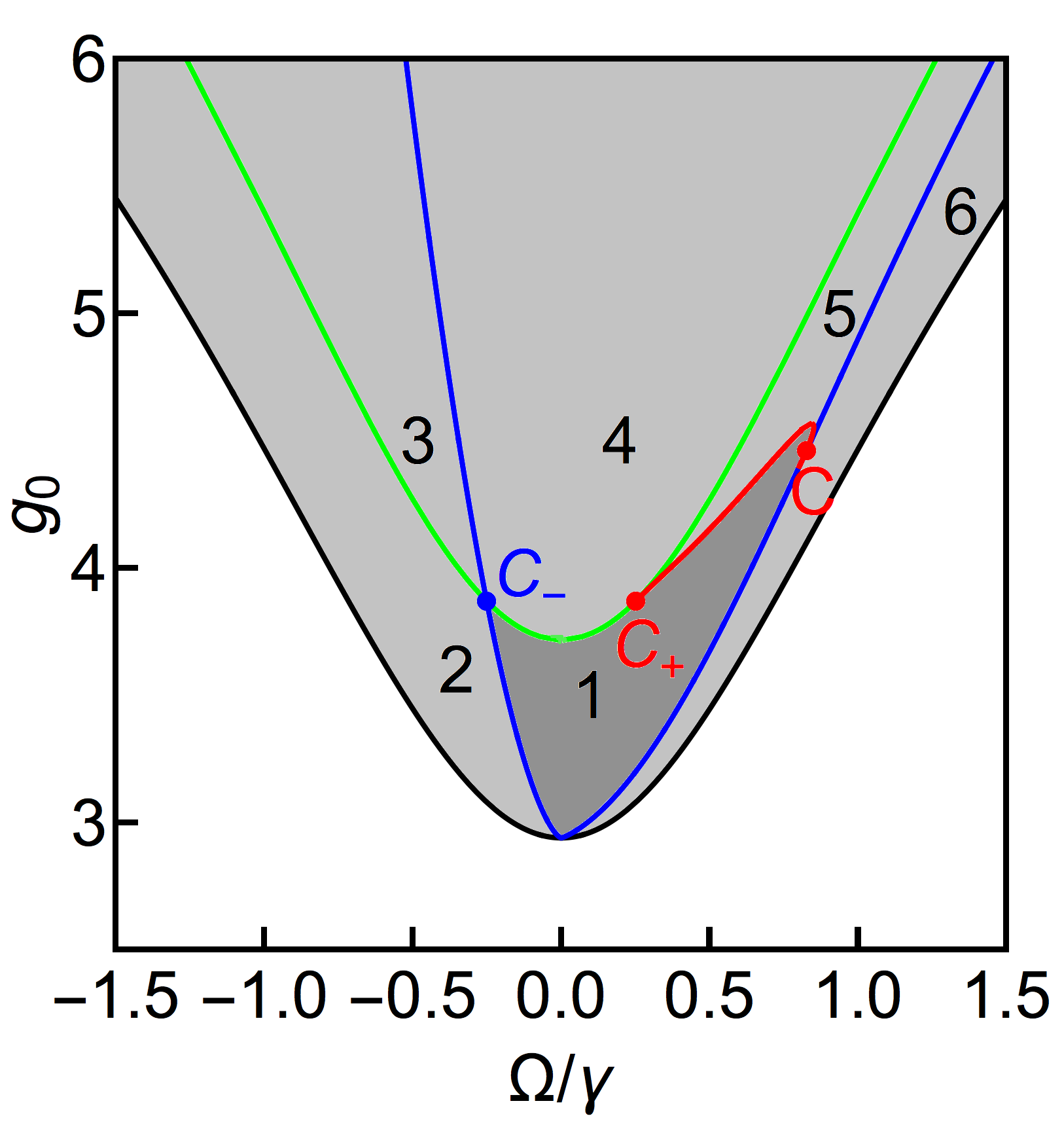}
\caption{Left: Two branches of pseudo-continuous spectrum. Different numbers illustrate qualitative behaviour of the branches in numbered parameter domains of the right panel.
Right: Bifurcation diagram  on the plane of two parameters: CW frequency offset $\Omega$ and pump parameter $g_0$, obtained in the limit of large delay time.
Black -- fold bifurcation curve defined by Eq.~\eqref{eq:SN}, blue -- modulational instability curve, which serves as a boundary of the Busse balloon and is defined by Eq.~\eqref{eq:modulational} in the Appendix, green -- flip instability leading to square wave appearance and defined by Eq.~\eqref{eq:SW}, red -- Turing-type (wave) instability. $C_{\pm}$  are codimension-two points defined by Eqs.~\eqref{eq:PD} and \eqref{eq:CTpm}. CW solutions are stable (unstable) in dark  gray region labeled ``1'' (light gray regions). $\gamma=2$ and $T\to\infty$. Other parameters are as for Fig.~\ref{fig:CWbranches}}\label{fig:stability}
\end{figure*}

To study linear stability the CW solutions of Eq.~\eqref{eq:model} we apply the large delay limit approach described in Ref.~\cite{Yanchuk2010a}. By linearizing Eq.~\eqref{eq:model} on a CW solution given by Eq.~\eqref{eq:CW} we obtain the characteristic equation in the form:
\begin{equation}
c_2Y^2+2c_1(\lambda)Y+c_0(\lambda)=0,\label{eq:characteristic}
\end{equation}
where $Y=e^{-\lambda T}$ and $c_{0,1,2}$ are given in the Appendix.  In the large delay limit, $\gamma T\gg 1$, the eigenvalues belonging to the so-called {\em pseudo-continuous spectrum} can be represented in the form:
$$\lambda= i \lambda_0 + \frac{\lambda_1}{T}+{\cal O}\left(\frac{1}{T^2}\right),\quad \lambda_1=\lambda_{11}+i\lambda_{12},$$ 
with real $\lambda_0$, $\lambda_{11}$, and $\lambda_{12}$. Therefore, using the approximate relations $c_{0,1}(\lambda)\approx c_{0,1}(i\lambda_0)$ and $Y\approx e^{-i \lambda_0 T-\lambda_1}$ we can solve the characteristic equation to express  $\lambda_{11}$ as a function of $\lambda_0$ \cite{Yanchuk2010a}
\begin{eqnarray}
\lambda_{11}^{\pm}=\Re{\lambda_1^{\pm}}=\Re{\ln\left(Y_{\pm}^{-1}\right)},\nonumber\\ Y_{\pm}=\frac{-c_1(i\lambda_0)\pm\sqrt{c_1(i\lambda_0)^2-c_0(i\lambda_0)c_2}}{c_2}.\label{eq:pseudo}
\end{eqnarray}
Two solutions $\lambda_{11}^{\pm}(\lambda_0)$ given by Eq.~\eqref{eq:pseudo} define two branches of pseudo-continuous spectrum shown in the left panel of Fig.~\ref{fig:stability}. Due to the phase shift symmetry of the model equation \eqref{eq:model} $E\to Ee^{i\varphi}$ with arbitrary constant $\varphi$ one of these branches is tangent to the $\lambda_{11}=0$ axis on the ($\lambda_0$,$\lambda_{11}$)-plane at the point $\lambda_0=0$, i.e., $Y_{-}^{-1}\rvert_{\lambda_0=0}=1$. The right panel of Fig.~\ref{fig:stability} shows the stability diagram of the CW solutions on the plane of two parameters: normalized frequency offset $\Omega/\gamma$ of a CW solution and pump rate $g_0$. CW solutions are stable (unstable) in the dark (light) gray domains.  Black curve indicates the fold bifurcation, where two  CW solutions merge and disappear.  Below this curve calculated from Eqs.~\eqref{eq:CW1} and \eqref{eq:SN} there are no CW solutions while above it a pair of CW solutions is born with one them corresponding to smaller intensity $R$ being always unstable. Similarly to the case of Eckhaus instability \cite{tuckerman1990bifurcation,wolfrum2006eckhaus} the upper branch of CW solutions can be stable only within the so-called Busse balloon (region labelled ``1'' in Fig.~\ref{fig:stability}) limited from below by modulational instability curve shown by blue line.  The modulational instability curve is defined by the condition that one of the two branches of the pseudo-continuous spectrum, which satisfies the relation $\left(\lambda_{11}^{-}\right)_{\lambda_0=0}=\left(d\lambda_{11}^{-}/d\lambda_0\right)_{\lambda_0=0}=0$,  changes the sign of its curvature at the point $\lambda_0=0$:
\begin{equation}
\left(\frac{d^2\lambda_{11}^{-}}{d\lambda_0^2}\right)_{\lambda_0=0}=0.\label{eq:MI}
\end{equation}
This instability is illustrated  sub-panels 2 and 6 of the left panel of Fig.~\ref{fig:stability}. Explicit expression for the modulational instability condition \eqref{eq:MI}  in terms of the model equation parameters is given in the Appendix.

 It is seen from the right panel of Fig.~\ref{fig:stability} that the modulational instability curve is tangent to the fold bifurcation curve at  $\Omega=0$ and that  this curve becomes asymmetric with respect to the axis $\Omega=0$ sufficiently far away from the tangency point.

The upper boundary of the CW stability domain  shown in the right panel of  Fig.~\ref{fig:stability} consists of two parts separated by codimension-two point C$_+$. The right part of this boundary lying between the points C$_+$ and C is indicated by red line and corresponds to the so-called Turing-type (wave) instability \cite{Yanchuk2010a}, where one of the two branches of pseudo-continuous spectrum cross the $\lambda_{11}=0$ axis at two symmetric points with $\lambda_0\neq 0$  i.e., $\lambda_{11}^+=0$, at $\lambda_0=\pm\lambda_0^*$ with $\lambda_0^*>0$, see sub-panel 5 of the left panel of Fig.~\ref{fig:stability}. The left part of the stability boundary lying between two symmetric codimension-two points C$_{\pm}$  corresponds to a flip instability leading to the appearance of square waves with the period close to $2T$. In the right panel of Fig.~\ref{fig:stability} the flip instability curve is shown by green line defined by the condition 
\begin{equation}
Y_+^{-1}\rvert_{\lambda_0=0}=-1, \label{eq:SW}
\end{equation} 
which can be rewritten in the form:
\begin{equation}
2-\frac{g_0 R}{(1+R)^2}+aR \cot\left(\frac{aR}{2}\right)=0.\label{eq:PD}
\end{equation}
The  codimension-two points C$_{\pm}$ are defined by Eq.~\eqref{eq:PD} together with additional  conditions $\left(d^2\lambda_{11}^{\pm}/d\lambda_0^2\right)_{\lambda_0=0}=0$. Using the relation \eqref{eq:PD} the additional conditions can be rewritten as 
\begin{equation}
\Omega^2\pm\gamma \Omega a R =\gamma^2.\label{eq:CTpm}
\end{equation}
An implicit equation for the coordinates of C$_{\pm}$ on the ($\Omega$,$g_0$)-plane are obtained by solving Eq.~\eqref{eq:CTpm} for $R$ and substituting the resulting solution into Eq.~\eqref{eq:PD}. It follows from Eq.~\eqref{eq:CTpm} that the codimension-two points C$_{\pm}$ shown in the right panel of Fig~\eqref{fig:stability} are symmetric with respect to $\Omega=0$ axis. It is seen from the right panel of  Fig.~\ref{fig:stability}  that the central longitudinal mode with $\Omega=0$ is the first one undergoing a flip instability with the increase of the pump parameter $g_0$. The larger is the frequency offset of the mode, the higher is the flip bifurcation threshold  for this mode. When, however, positive (negative) frequency offset is sufficiently large the mode is already unstable with respect to  Turing (modulational) instability at the flip instability point. In this case the flip bifurcation results in the appearance of unstable square waves. 
Note, that for any finite values of delay time, $\gamma T<\infty$, it is an Andronov-Hopf bifurcation of a CW solution that leads to the emergence of square waves. However, in the limit of infinite delay the period of square wave regime tends to infinity, which means that the imaginary parts of complex eigenvalues crossing the imaginary axis at the Andronov-Hopf bifurcation tend to zero. Hence, in the limit of infinite delay we refer to this bifurcation as a flip instability.

Let us consider  the  CW solution  of Eq.~\eqref{eq:model} corresponding to the central longitudinal mode with zero detuning from the central frequency  of the spectral filter, $\Omega=0$. For this solution the two quantities $Y_{\pm}$ in Eq.~\eqref{eq:pseudo} take the form
\begin{eqnarray}
Y_{-}\rvert_{\Omega=0}&=&1-i\frac{\lambda_0}{\gamma},\label{eq:Ym}\\
Y_{+}\rvert_{\Omega=0}&=&\frac{1-i \lambda_0/\gamma }{1+a R \cot {\left(\frac{a R}{2}\right)}-\frac{g_0 R}{(1+R)^2}}.\label{eq:Yp}
\end{eqnarray}
From Eq.~\eqref{eq:Ym} we get the relations $Y_{-}^{-1}\rvert_{\Omega=0,\lambda_0=0}=1$ and $\left|Y_{-}^{-1}\right|_{\Omega=0,\lambda_0\neq0}<1$ meaning that the first branch of the pseudo-continuous spectrum of the central longitudinal mode is always stable and is tangent to the imaginary axis at $\lambda_0=0$.  On the other hand, from Eq.~\eqref{eq:Yp} we see that the fold bifurcation of the central longitudinal mode  defined by the condition $Y_{+}^{-1}\rvert_{\Omega=0,\lambda_0=0}=1$ coincides with Eq.~\eqref{eq:SN}. Similarly the flip instability responsible for the emergence of square-waves is defined by the condition $Y_{+}^{-1}\rvert_{\Omega=0,\lambda_0=0}=-1$ coinciding with Eq.~\eqref{eq:PD}.
As it will be shown in the next section,  the  condition \eqref{eq:PD} defines also the period doubling bifurcation of a 1D map, which we construct in the next section to study the square wave formation in the DDE model Eq.~\eqref{eq:model}. 

\begin{figure*}
\includegraphics[width=0.45\textwidth]{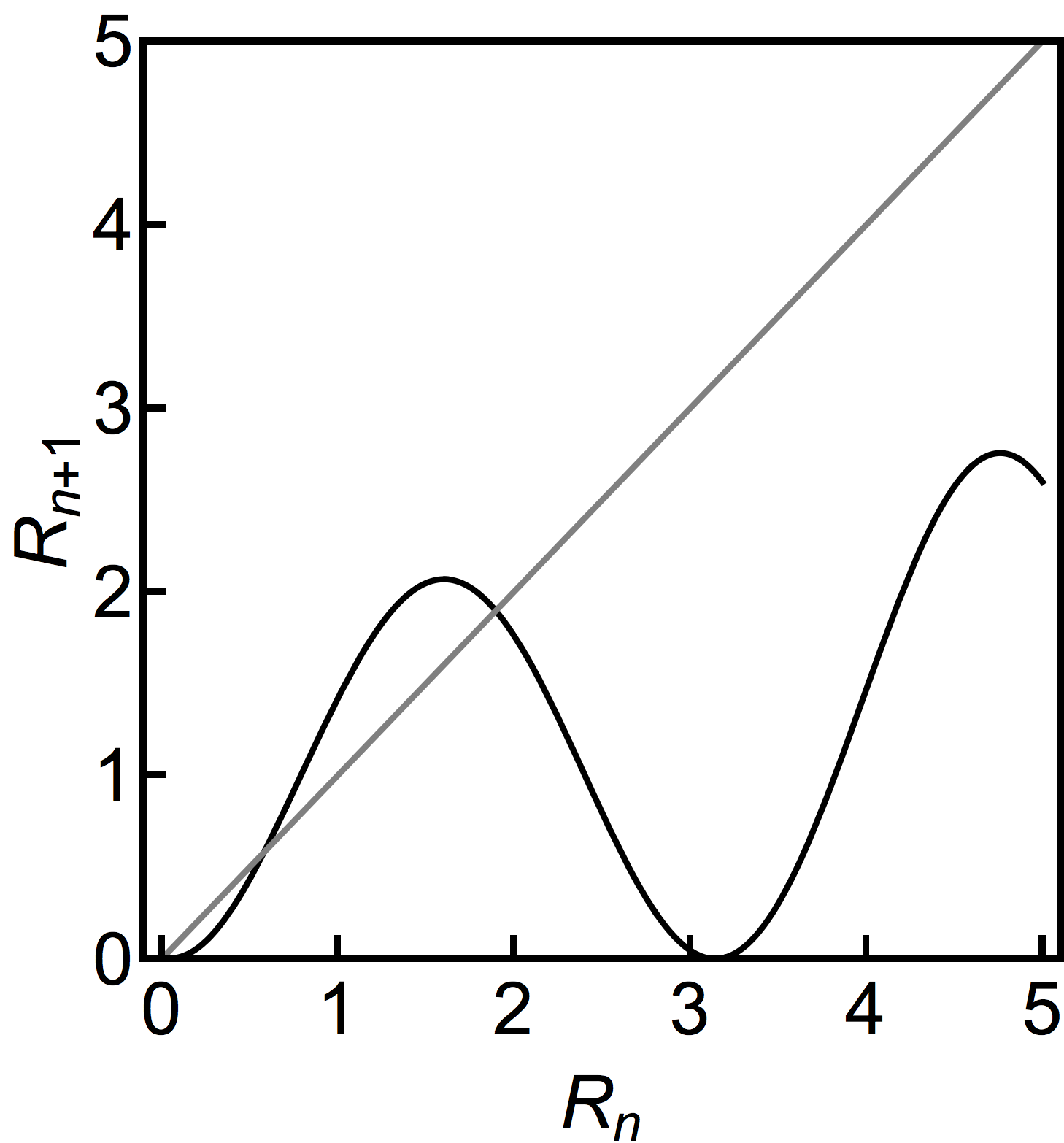}\quad
\includegraphics[width=0.47\textwidth]{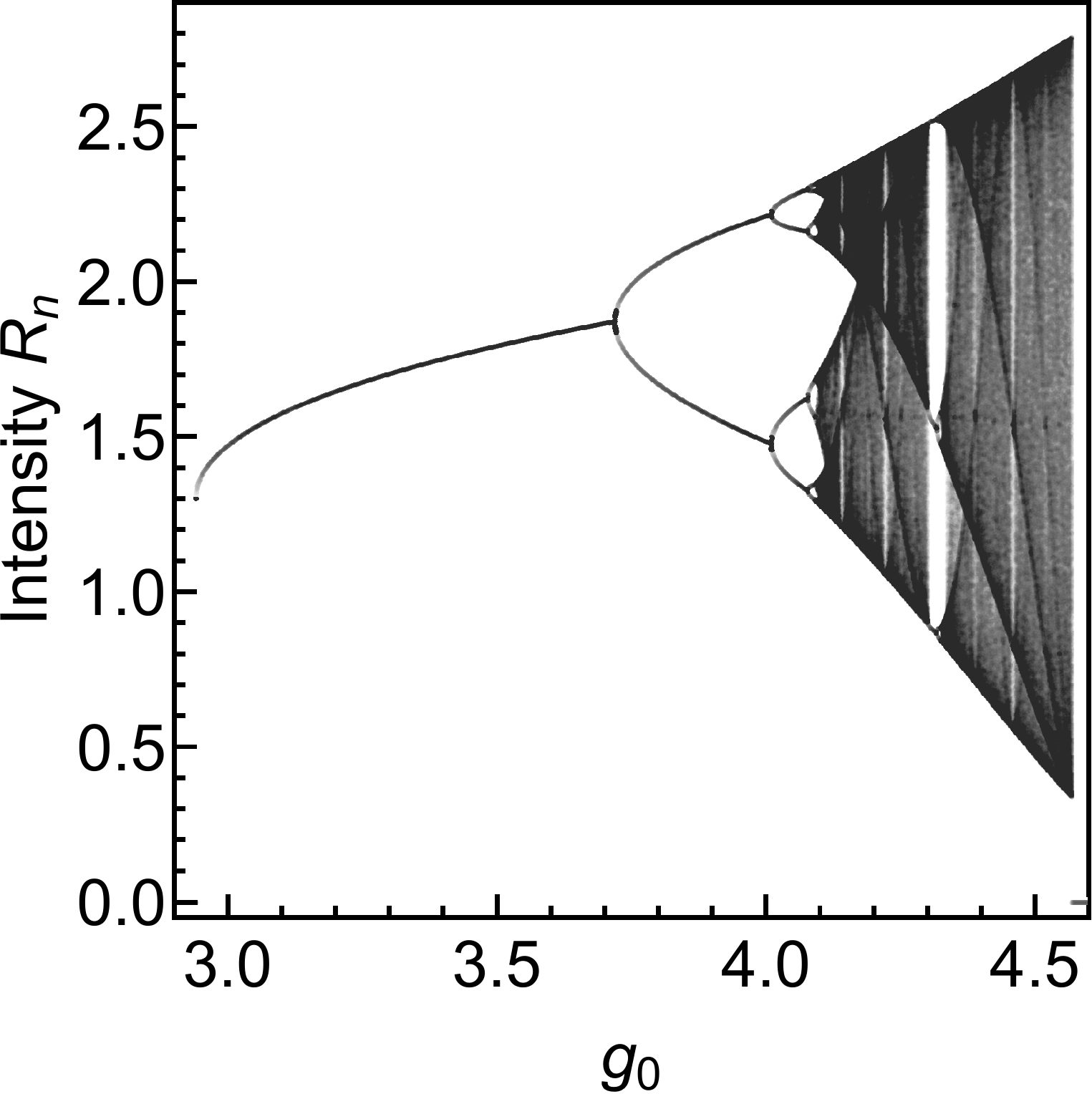}
\caption{Left: Graph of the function $h$ defined by Eq.~\eqref{eq:1dmap}. Two period one fixed points of the map correspond to the intersections of the black curve with straight gray line $R_{n+1}=R_n$. These fixed pointa correspond to the CW solutions of Eq.~\eqref{eq:model} lying on the upper and lower part of the CW branch with zero frequency offset $\Omega=0$. $a=2$, $\kappa=0.3$, and $g_0=3.8$. Right: Bifurcation diagram illustrating period-doubling route to chaos  in the map \eqref{eq:1dmap} with $a=2$ and $\kappa=0.3$. }\label{fig:doubling}
\end{figure*}

\section{1D map and square waves}
The existence of stable square wave solutions in Eq.~\eqref{eq:model} can be demonstrated by constructing a 1D map \cite{chow1983singularly} that exhibits a period doubling bifurcation corresponding to the emergence of square waves in the DDE model \eqref{eq:model}. To this end we rescale the time $\tau=t/T$  in Eq.~\eqref{eq:model} and obtain
\begin{equation}
\varepsilon\frac{dE}{d\tau}+E=\sqrt{\kappa}e^{g_{0}/[2(1+|E(\tau-1)|^{2})]}f\left[|E(\tau-1)|^{2}\right]E(\tau-1),
\label{eq:model1}
\end{equation}
where in the large delay limit we have $\varepsilon\equiv 1/\gamma T\ll 1$. By discarding the time derivative term, which is proportional to the small parameter $\varepsilon$, we transform this equation into a complex map, which describes the transformation of the electric field envelope $E$ after a round trip in the cavity and resembles the well known Ikeda map \cite{ikeda1979multiple}. Then, taking modulus square of both sides we obtain a 1D map for the intensity $R$:
\begin{equation}\label{eq:1dmap}
R_{n+1}=h(R_n),\quad h(R_n)=\kappa e^{\frac{g_0}{1+R_n}}\sin^2\left(\frac{aR_n}{2}\right)R_n,
\end{equation}
where $R_n\equiv R(n)$ ($n=0,1,2\dots$) with fixed points $R_n=R^*$ satisfying the condition $R^*=h(R^*)$:
\begin{equation}
\kappa e^{\frac{g_{0}}{1+R^*}}\sin^{2}\left(\frac{aR^*}{2}\right)=1.\label{eq:fixedpoint}
\end{equation}
Graphical representation of the function $h$ is given in the left panel of Fig.~\ref{fig:doubling}. Note that since Eq.~\eqref{eq:fixedpoint} is equivalent to  Eq.~\eqref{eq:CW1}  taken at $\Omega=0$ and $\Omega=0$ is a solution of Eq.~\eqref{eq:CW2}, fixed points of the map \eqref{eq:1dmap} have the intensity $R$ coinciding with that of the central longitudinal mode, i.e. the CW solution of Eq.~\eqref{eq:model} with zero frequency offset $\Omega=0$  from the central frequency of the spectral filter. Furthermore, for sufficiently large $g_0$ a stable fixed point of the map \eqref{eq:1dmap} exhibits a period doubling bifurcation which is defined by
\begin{equation}
1-\frac{g_0 R^*}{(1+R^*)^2}+aR^* \cot\left(\frac{aR^*}{2}\right)=-1, \label{eq:PDmap}
\end{equation}
together with \eqref{eq:fixedpoint}. Since the relations \eqref{eq:fixedpoint} and \eqref{eq:PDmap} are equivalent to \eqref{eq:CW1} and \eqref{eq:SW} evaluated at $\Omega=0$, the period doubling bifurcation point of the map \eqref{eq:1dmap} coincides in the limit of large delay with the flip instability to square waves of the central longitudinal mode having zero frequency offset, $\Omega=0$. 
It is seen from right panel of Fig.~\ref{fig:doubling} that after the first period doubling bifurcation  the 1D map \eqref{eq:1dmap} demonstrates a period doubling transition to chaos.

\begin{figure*}
\begin{centering}
\includegraphics[width=0.45\textwidth]{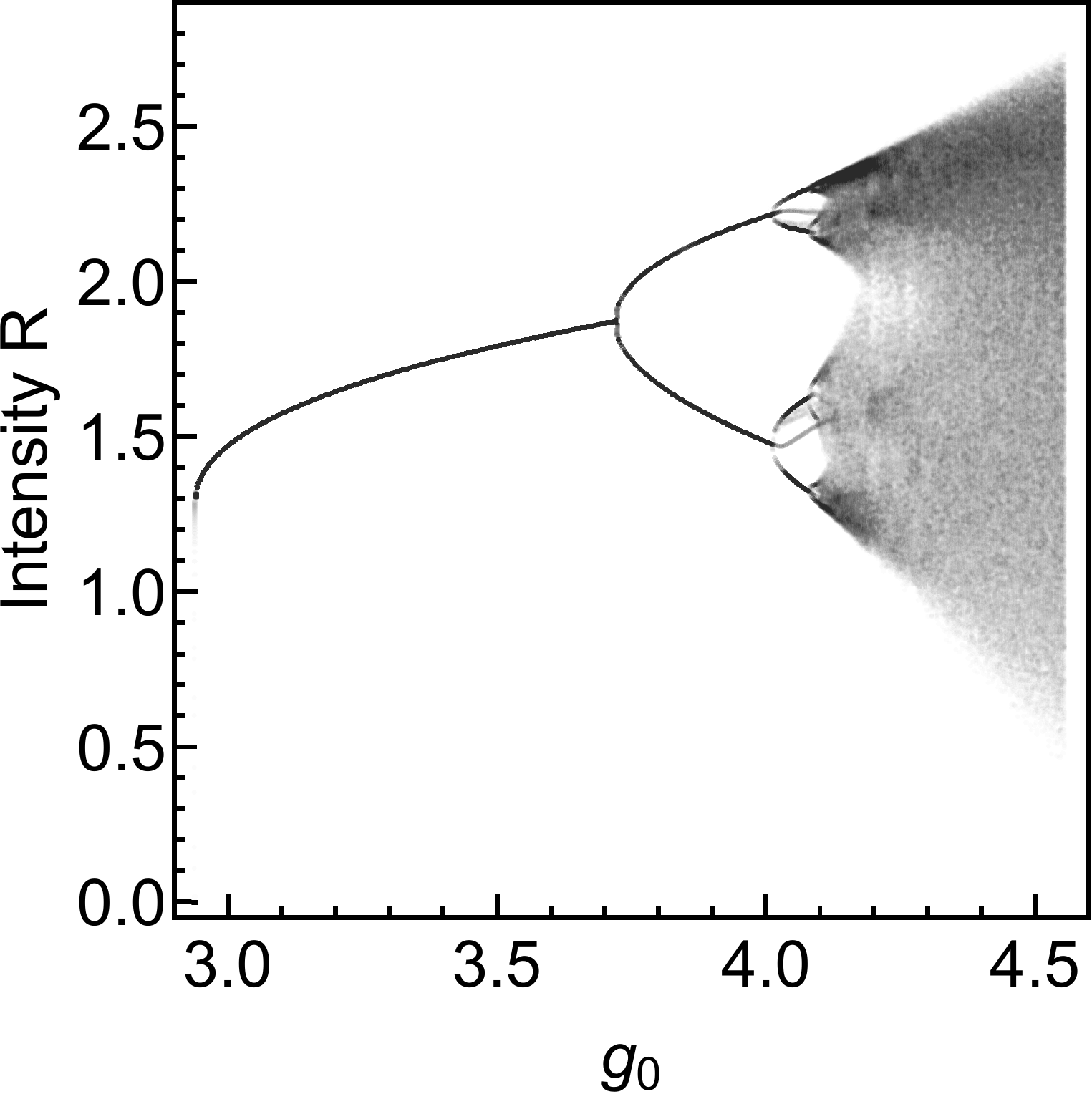}\quad
\includegraphics[width=0.48\textwidth]{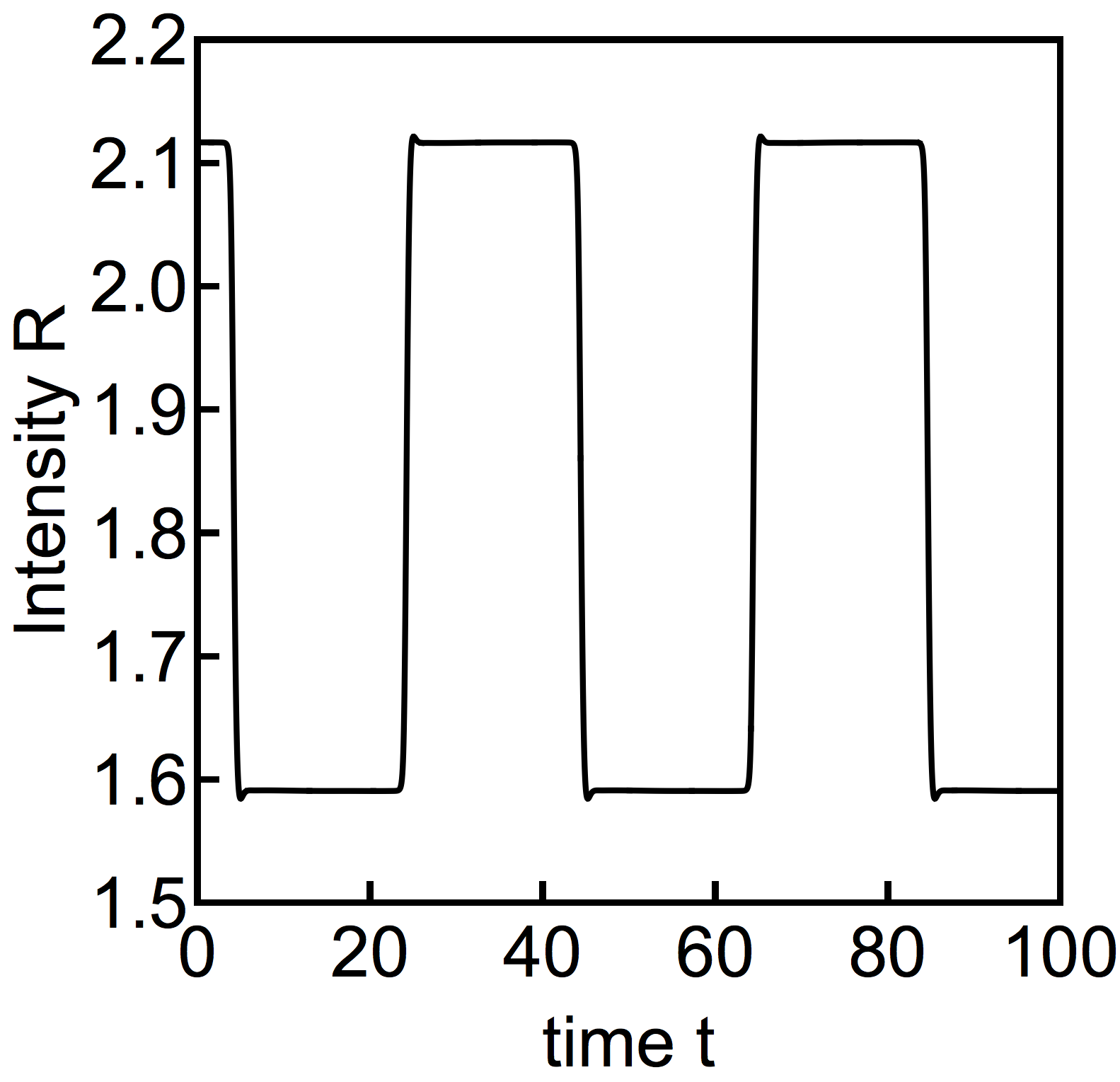}
\end{centering}
\caption{Left: Bifurcation diagram obtained by numerical integration of Eq. (\ref{eq:model})
with $\kappa=0.3$, $a=2$, $T=20$, and $\gamma=5$. Right: Square waves calculated numerically for $g_{0}=4.0$. \label{fig:squarewaves}}
\end{figure*}

Period-doubling route to chaos obtained by numerical integration of the DDE model \eqref{eq:model} is illustrated in left panel of  Fig.~\ref{fig:squarewaves}, where local maxima of the electric field intensity time-trace are plotted versus increasing values of the pump parameter $g_0$. It is seen that the diagram in this figure is very similar to that obtained with the 1D map \eqref{eq:1dmap}, cf. Fig.~\ref{fig:doubling}. Note, however, that the period doubling threshold is slightly higher in Fig.~\ref{fig:squarewaves}  than in Fig.~\ref{fig:doubling}. This can be explained by taking into consideration that in the DDE model \eqref{eq:model} the threshold of the flip instability leading to the emergence of square waves increases with the absolute value of the frequency detuning $\Omega$. Therefore, we can conclude that in Fig.~\ref{fig:squarewaves} the CW solution undergoing the period-doubling cascade must have a small non-zero frequency detuning, $\Omega\neq 0$.  The first period doubling bifurcation in the left panel of  Fig.~\ref{fig:doubling} is responsible for the formation of square waves shown in the right panel of Fig.~\ref{fig:doubling}, while further period doublings give rise to more complicated square wave patterns with larger periods. Finally, we note that with the increase of the gain parameter $g_0$ new pairs of fixed points of the map \eqref{eq:1dmap} appear in saddle-node bifurcations. For example, the second pair of fixed points corresponds to the second (right) maximum of the function $h$ shown in the left panel of Fig.~\ref{fig:doubling} and to additional branches of CW solutions visible in the upper right part of the right panel of Fig.~\ref{fig:CWbranches}. 
Although the linear stability analysis performed in the previous section is applicable to these additional high intensity CW solutions as well, here we restrict our consideration to moderate pumping levels.

\begin{figure*}
\includegraphics[width=0.45\textwidth]{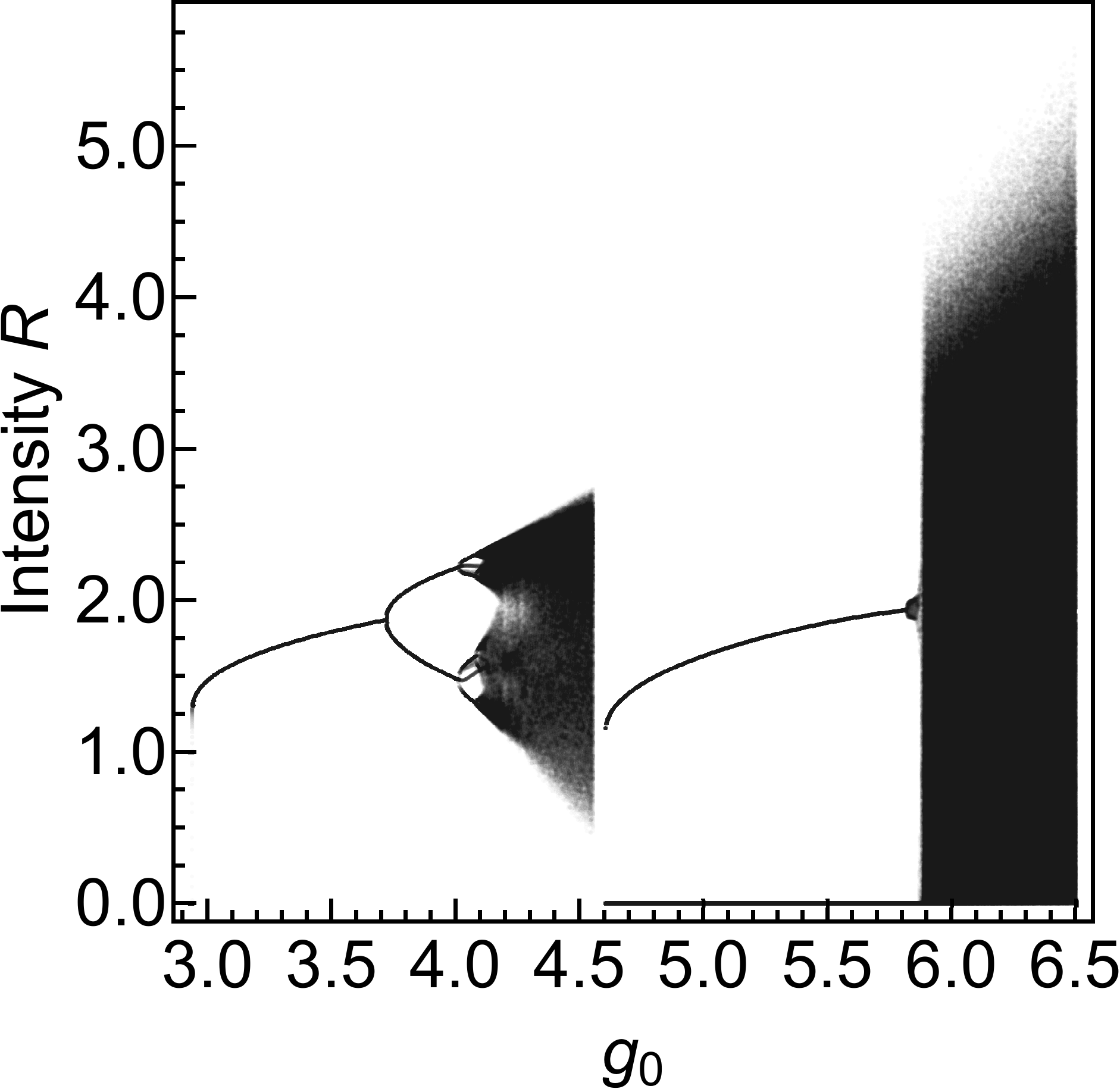}
\includegraphics[width=0.46\textwidth]{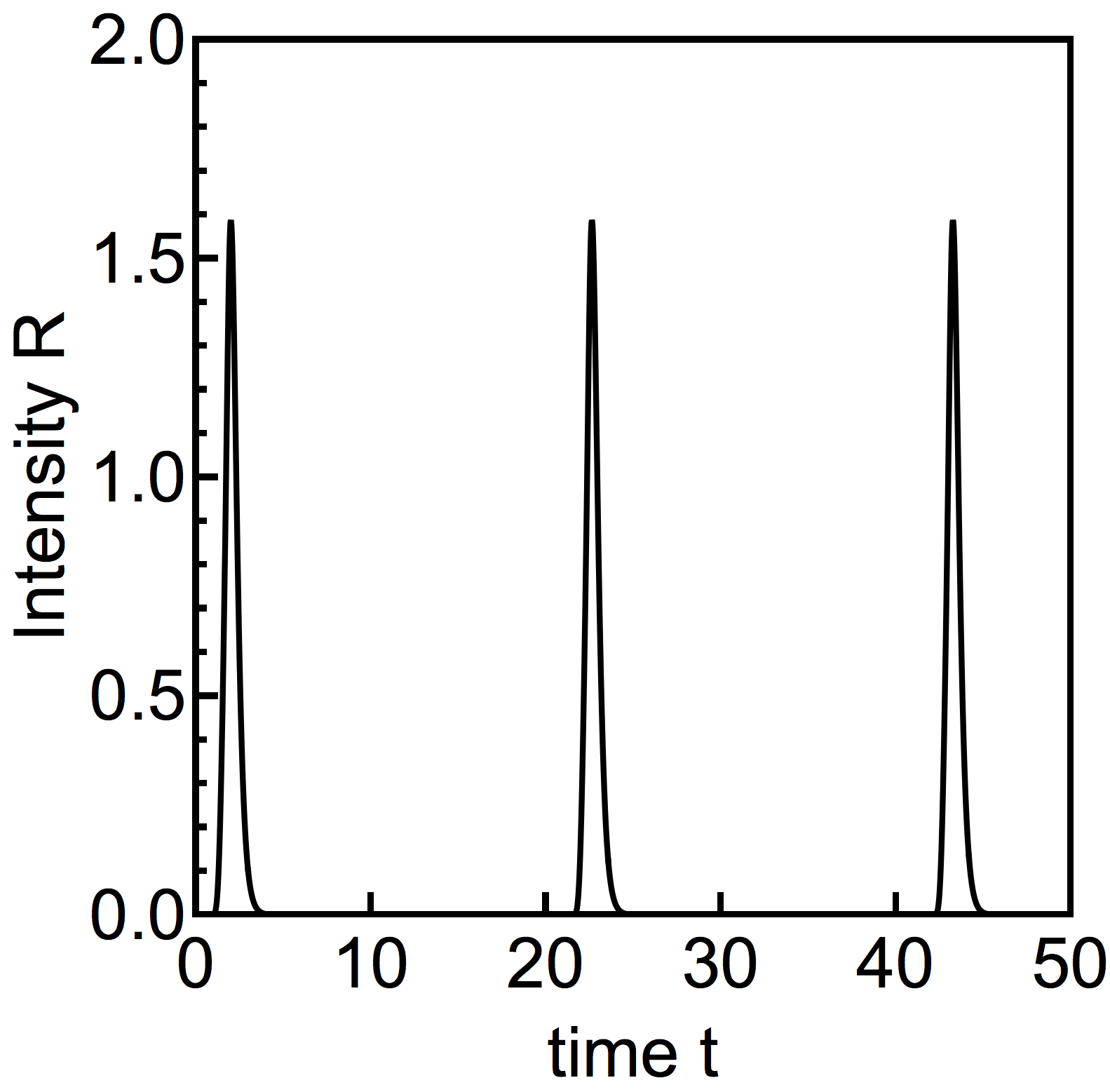}
\caption{Left: Bifurcation diagram obtained by numerical integration of Eq. (\ref{eq:model}). Right: fundamental mode-locked regime with the repetition period close to the cavity round trip time calculated for $g_0=5.0$.  Other parameters are same as for Fig. \ref{fig:squarewaves}. }
\label{fig:full}
\end{figure*}

\section{Mode-locking}
Bifurcation diagram in the left panel of Fig.~\ref{fig:full} was obtained by numerical integraton of Eq.~\eqref{eq:model}. It is similar to that  shown in  Fig.~\ref{fig:squarewaves}, but  spans a larger range of the pump parameter values. It follows from this diagram that with the increase of the pump parameter  $g_0$ after a chaotic regime associated with the period doubling cascade the phase trajectory of the system  jumps to a pulsed solution with time periodic laser intensity. This solution corresponding to a fundamental mode-locked regime with the repetition period close to the cavity round trip time, $T$, is illustrated in right panel of Fig.~\ref{fig:full}, where it is seen that the leading edge the pulses is much steeper than the trailing edge. In the following, we will show that unlike the trailing tail of the pulses, which decays exponentially,  their leading tail demonstrates faster-than-exponential (super-exponential) decay in negative time.

Let us consider a mode-locked solution of Eq.~(\ref{eq:model}) with the period $T_{0}$ close to the cavity round trip time $T$. For this solution satisfying the condition $E(t)=E(t+T_{0})e^{i\Delta}$ we can write
\begin{equation}
E(t-T)=E(t-T+T_{0})e^{i\Delta}\equiv E(t+\delta)e^{i\Delta}\label{eq:periodic}
\end{equation}
where $\delta=T_{0}-T>0$ is the small difference between the solution period and the delay time. Substituting  \eqref{eq:periodic} into \eqref{eq:model} we get a time advance equation
\begin{equation}
\gamma^{-1}\frac{dE}{dt}+E=\frac{\sqrt{\kappa}}{2}e^{g_{0}/[2(1+|E(t+\delta)|^{2})]+i\Theta}f(|E(t+\delta)|^{2})E(t+\delta),\label{eq:delta}
\end{equation}
where $\Theta=\theta+\Delta$. Note, that unlike the original DDE model \eqref{eq:model}, which has stable trivial solution $E=0$, the trivial solution of Eq.~\eqref{eq:delta} is a saddle with one stable and an infinite number of unstable directions.  The stable direction determines the decay rate of the trailing tail of mode-locked pulses,  while unstable directions are responsible for the decay of the leading tail in negative time.  To perform linear stability analysis of the trivial solution of Eq.~\eqref{eq:delta} we write the following equation linearized at $E=0$:
\begin{equation}
\gamma^{-1}\frac{dE}{dt}+E=\epsilon E(t+\delta)e^{i\Theta},
\label{eq:lin}
\end{equation}
where linear time advance term in the right hand side proportional to small perturbation parameter $\epsilon$ describes an imperfection introduced by a slight asymmetry of the  coupler between the laser cavity and the NALM.
The spectrum of Eq.~(\ref{eq:lin}) is defined by
\begin{equation}
\lambda_{k}=-\gamma\left[1+\frac{W_{k}(-\epsilon\gamma\delta e^{-\gamma\delta+i\Theta})}{\gamma\delta}\right],
\end{equation}
where $W_{k}$ denotes the $k$th branch of multivalued Lambert function. In particular, in
the limit $\epsilon\to0$ the eigenvalue with the index $k=0$ is
the only stable and negative one $\lambda_{0}\to-\gamma<0$. This eigenvalue determines the decay rate of the trailing tail of mode-locked pulses. The remaining eigenvalues with $k\neq 0$ have positive real parts diverging in the limit $\epsilon\to0$, $\Re\lambda_{k}\to+\infty$. Among these unstable eigenvalues, depending on the value of $\Theta$, one of the two eigenvalues $\lambda_{\pm 1}$ with $k=\pm 1$ has the smallest real 
part. All other eigenvalues with $|k|>1$  have larger real parts. 
Hence,  for small nonzero $\epsilon$ generically one of the two eigenvalues $\lambda_{\pm 1}$ determines the decay rate  the pulse leading tail in negative time. The fact that this eigenvalue tends to infinity as $\epsilon\to0$ suggests that this decay is super-exponential. 

\begin{figure}
\begin{centering}
\includegraphics[scale=0.6]{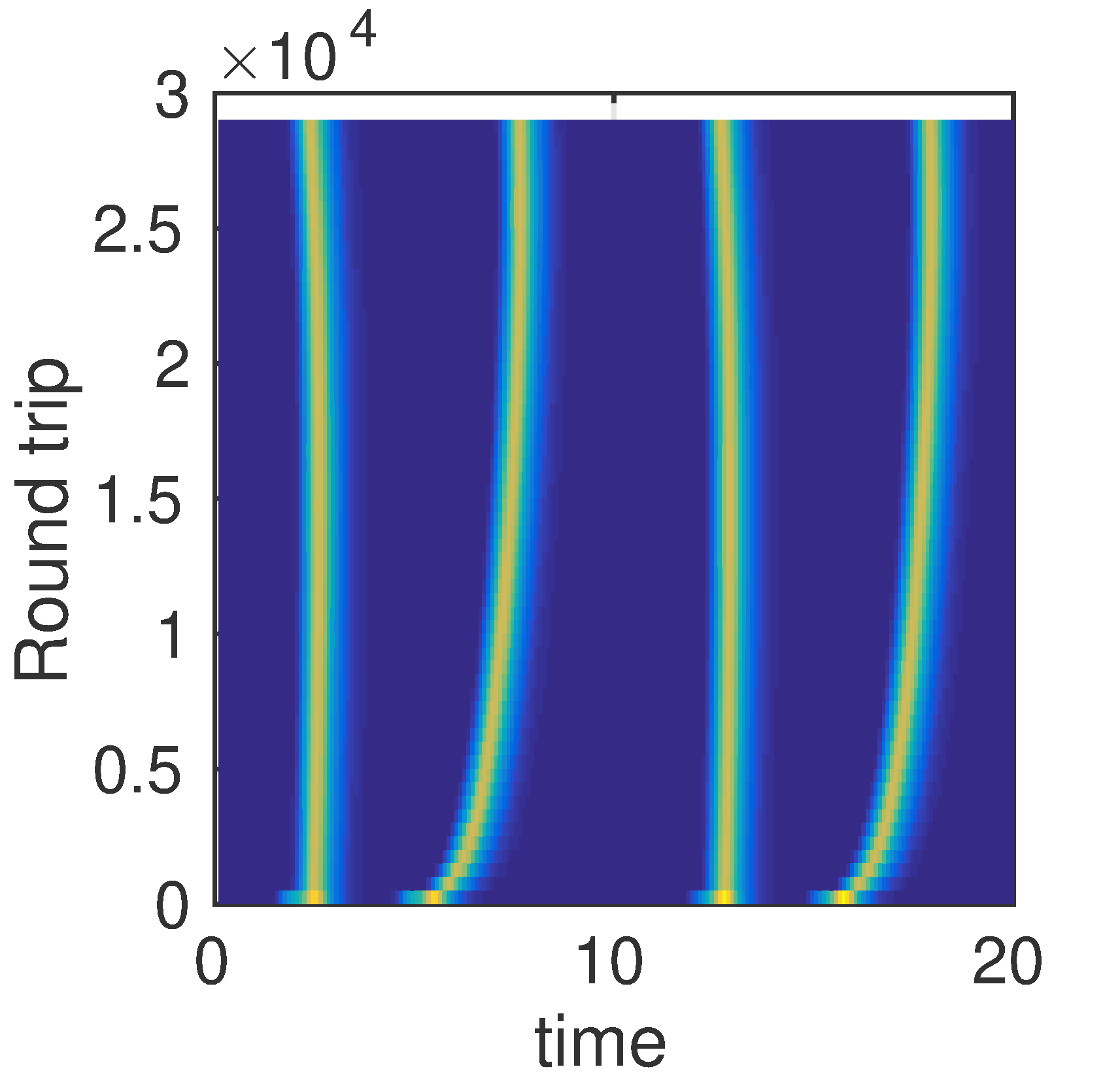}
\caption{Interaction of two mode-locked pulses leading to a harmonic mode-locked regime with two pulses per cavity round trip. A common drift of the two interacting pulses is eliminated. $T=10$. The time axis spans the interval $2T$. Other parameters are same as for Fig.~\ref{fig:squarewaves}.  }
\label{fig:interaction}
\end{centering}
\end{figure}

Finally let us discuss briefly the interaction of asymmetric mode-locked pulses shown in the right panel of Fig.~\ref{fig:full}. When integrating the model equation \eqref{eq:model} numerrically it is possible to seed  two or more non-equidistant pulses in the laser cavity as an initial condition. Then the pulses start to interact locally via their decaying tails. The asymmetric nature of these tails suggests that similarly to the case discussed in \cite{vladimirov2018effect} local interaction of non-equidistant pulses will be very asymmetric as well.  This can be is seen in Fig.~\ref{fig:interaction} illustrating an interaction of two asymmetric pulses of Eqs.~\eqref{eq:model} on the time-round trip number plane.  We see that two initially non-equidistant pulses repel each other and tend to be equidistantly spaced in the long time limit. Furthermore, when the two pulses are sufficiently close to one another, exponentially decaying trailing tail of the left pulse repels noticeably the right pulse, while super-exponentially decaying leading tail of the right pulse almost does not affect the position of the left pulse.  When the pulses become equidistant the forces acting on a pulse from opposite directions balance each other and a stable harmonic mode-locking regime with two pulses per cavity round trip time is established. 

\section{Conclusion}
We have considered a simple  DDE model of unidirectional class-A ring NALM mode-locked laser.  Linear stability analysis in the  large delay limit revealed that similarly to the well-known Eckhaus instability only those CW solutions, which belong to the Busse balloon can be stable. This balloon is limited from below by the modulational instability boundary.  We demonstrated that with the increase of the pump parameter CW regimes loose their stability either via a Turing-type instability, or through a flip instability leading to a formation of stable square waves.  We have constructed a 1D map which describes the transition to square waves and their secondary bifurcations giving rise to a more complicated square wave patterns with larger and larger periods. We have shown that mode-locked pulses, which appear  after a chaotic square-wave dynamics, have exponentially decaying trailing tail and a leading tail, which decays super-exponentially in negative time.  When two or more pulses circulate in the laser cavity, the interaction of these pulses is repulsing and very asymmetric leading to a formation of harmonic mode-locked regimes. Noteworthy, that the mode-locked regime considered here is always non-self-starting, which means that the pulses are sitting on a stable laser-off solution. Hence, these pulses can be viewed as temporal cavity solitons having similar properties to spatial and temporal localized structures of light observed in bistable optical systems. 

\bigskip

Authors thank D. Turaev for useful discussions. A.V.K. and E.A.V. acknowledge the support by Government of Russian Federation (Grant 08-08). A.G.V. acknowledges the support of the F{\'e}d{\'e}ration Doeblin CNRS and SFB 787 of the DFG, project B5.

\section{Appendix}
The coefficients $c_0(\lambda)$, $c_1(\lambda)$, and $c_2$ in the characteristic equation \eqref{eq:characteristic} are defined by
\begin{widetext}
\begin{eqnarray}
c_0(\lambda)&=&(\gamma -\lambda )^2+\Omega ^2,\nonumber\\
c_1(\lambda)&=&-\frac{a \lambda  R (R+1)^2 \Omega +\gamma  (\gamma -\lambda ) [R (2 R+4-g_0)+2]+\Omega ^2 [R (2
   R+4-g_0)+2]}{(R+1)^2}-a R \cot \left(\frac{a R}{2}\right) \left(\gamma ^2-\gamma  \lambda +\Omega ^2\right),\nonumber\\
c_2&=&\frac{\gamma ^2 \kappa  e^{\frac{g_0}{R+1}}}{2
   (R+1)^2} \{a R (R+1)^2 \sin (a R)-[R (R+2-g_0)+1] [\cos (a R)-1]\}.\nonumber
\end{eqnarray}
The modulational instability condition \eqref{eq:MI} of the CW solutions of Eq.~\eqref{eq:model} can be rewritten in the following form:
\begin{eqnarray}
\frac{\left[a(1+R)^{2}\Omega+\gamma(\tilde g-g_{0})\right]^{2}}{\left(\gamma^{2}+\Omega^{2}\right)^{2}}+\frac{2(1+R)^{2}\Omega^{2}\left[ g_0^2+2a^2(1+R)^4-g_0^2 \cos{(aR)}-2ag_0(1+R)^2\sin{(aR)}\right] }{R(\tilde g-g_{0})\left(\gamma^{2}+\Omega^{2}\right)^{2}\left[1-\cos\left(aR\right)\right]}=0.
\label{eq:modulational}
\end{eqnarray}
\end{widetext}

%

\end{document}